\documentstyle[russian]{article}
\begin{document}
JETP Lett., Vol. 64, No. 6, 25 Sept. \copyright 1996 

\begin{center}

{\bf\large  Scale invariance in percolation theory and fractals.}
\\ \medskip {\it M.V. Entin, G.M. Entin.}\\ {\it \small
Institute of Semiconductor Physics, Siberian Branch of the
Russian Academy of Sciences,\\ 630090 Novosibirsk, Russia.\\
Novosibirsk State University, 630090 Novosibirsk, Russia
\\E-mail:Entin@isp.nsc.ru\\ }

\end{center}


\begin{center}
\parbox[c]{13.5cm}{
\normalsize

 The properties of the similarity transformation in percolation
theory in the complex plane of the percolation probability are
studied. It is shown that the percolation problem on a
two-dimensional square lattice reduces to the Mandelbrot
transformation, leading to a fractal behavior of the percolation
probability in the complex plane. The hierarchical chains of
impedances, reducing to a nonlinear mapping of the impedance space
onto itself, are studied. An infinite continuation of the procedure
leads to a fixed point. It is shown that the number of steps
required to reach a neighborhood of this point has a fractal
distribution. }\\ \end{center}

\large
\medskip

In percolation theory, an approach based on a hierarchical chain of
resistances is used to calculate the percolation threshold and the
behavior of a system near the percolation threshold. The approach
is based on a recurrence relation for the percolation probability
$p$ \cite{1} of a medium or for the resistances of the
hierarchically constructed cluster (see, for example, Ref. 2).

A very simple model of a scale transformation for geometric
percolation employs a calculation of the percolation probability
for a d-dimensional cube with edge length 2a and know percolation
probability a. If one neglects percolation between the diagonally
positioned conducting cubes, the renormalization-group relation for
the percolation probability at the nth level has the form \cite{1}:

\begin{eqnarray}
\label{1} p_n=1-(1-p_{n-1}^2)^{2^{d-1}}.
\end{eqnarray}

The mapping (1) is a particular case of more general formulations
of the renormaliztion-group study of percolation transitions. Our
objective in the present letter is to investigate the mapping (1)
in the complex $p$-plane. This mapping is nonlinear and its general
properties are described by the theory of fractals \cite{2}.
Another problem is to study a similar mapping for hierarchically
constructed chains of impedances simulating the high-frequency
properties of a percolation medium.

 Let us consider the two-dimensional (2D) case. The mapping (1)
possesses three fixed points: 0,1, and $p_á = (\sqrt{5}-1)/2$. The point
$p_c$ is unstable: for $p_0>p_c$, $p_n$ asymptotically approaches
1, and in the opposite case $p_n$ approaches zero. Therefore $p_c$
is the percolation threshold. The investigation of fractals is
based on the study of the very simple nonlinear Madelbrot map

\begin{eqnarray}
\label{2}
x'=F(x)=x^2+c.
\end{eqnarray}

 It is easy to see that choosing $c=-1$ , substituting Eq. (2) into
itself, and replacing $x$ by $-p$ gives precisely the map (1) with
$d=2$.  Therefore the Sarychev map (1) is a particular case of the
Mandelbrot map. By studying this map int the complex $p$-plane we
can determine the regions from which a conducting $(p_n\to 1)$ or
nonconducting $(p_n\to 0)$ medium is reached as $n\to \infty$. In terms of
the Mandelbrot theory, they are called basins of attraction of the
corresponding points

 The attraction basin of the point zero, corresponding to
nonpercolation in the 2D case, is displayed on Fig. 1. It is
distinguished by coloring the plane as a function of the parity of
number of steps required to reach a circle with radius $3 10^{-2}$
around the origin. The basins of attraction of other finite fixed
points are colored in black. The basin of attraction for zero cuts
off $(0,p_c)$ of the real axis the part from the segment  $(0,1)$
corresponding to the physical region. The percolation system is
determined by the segment $(p_c,1)$, which falls in the basin of
attraction of the point 1 - the black region on the right-hand
side.

 The fractal structure of the basin of attraction is formed by
regions spanning the points that map into zero after many
repetitions of the mapping, i.e. , the points are zeros of the
functions $F_{n}(p)=F(F(...F(p))...)$. This  function determines the
percolation probability of a square with an edge length $2^n$. The
derivative of $F_{n}(p)$ with respect to $p$ can be regarded as the
distribution function of the percolation thresholds. $F_n$ is a
polynomial of degree $2^n$ with even powers.

It is easy to obtain the asymptotic behavior of $F_n$ as
$\delta p=p- p_c \to 0$. Linearizing the map (2) gives
$\delta p'=2p_c \delta p $ i.e. $F_{n}'(p_c)\simeq (2 p_c)^n$.
Since $2p_c>1$, the derivative grows exponentially with $n$ and
$F_n(p)\to \Theta(p-p_c)$ as $n\to \infty$.

The general form of this function is, however, quite complicated.
It can be shown that the general formula for the zeros of this
polynomial has the form

\begin{eqnarray}
\label{3}
\sigma_1 \sqrt{1+\sigma_2
\sqrt{1+\sigma_3 \sqrt{....+\sigma_{n}}}}, \end{eqnarray}

where $\sigma_i=\pm 1$. Evidently, all zeros before the mappings are
also zeros after the mappings. The preceding zeros appear with a
higher multiplicity. Specifically, if $n = 2^k$, then the zero of
the $2^j$th mapping ($j < k$) appears with multiplicity $2^{k-j}$.

In the limit $n \to \infty$ the zeros approach the boundary of the basin of
attraction, forming a so-called Julia set. The fractal nature of
the Julia set can be discovered by examining the periodic
subsequences $\{ \sigma_i \}$. Each such subsequence approaches its own
irrational limit. Together the limits all form a fractal boundary.

It is easy to obtain a physical interpretation of a basin of attraction.
Consider the quantity $Re(\ln F_n (x) )$. This quantity is 2D
electrostatic potential of charges which are placed at zeros of the
function $F_n (x)$ and have a magnitude equal to the multiplicity
of the zeros. The equipotential lines form the basin of attraction.
Large, circular regions of influence of the charges arise near the
largest charges-the zeros of the mappings with small $j$. As $j$
increases, the number of regions increases as the number of zeros
and size $\rho_n$ of regions decreases. The rate of decrease of the
size can be found by equating the potential produced by the charges
at the points where the boundaries of the corresponding regions
touch:
$$2^{-j}\ln{\rho _{2^j}}=2^{-(j+1)} \ln{\rho _{2^{-(j+1)}}}$$

whence $\rho_n=\rho_0^n$. The choice of the set $ \{ \sigma_n \} $
equal to 1 gives, in accordance with the Lee-Yang theorem, a chain
of zeros of a polynomial which approach $p_c$.

Therefore the percolation probability behaves very simply on the axis in
the limit of large $n$, viz. as $\Theta(p-p_c)$, and in the complex plane
it exhibits a complicated, fractal behavior. If one intersted only the
asymptotic behavior of $F_n(p)$ for large $n$ and near a transition, then
there is no need to know the singularities of this function in the complex
plane. However, if we wish to describe the function for all $n$ and far
from a transition point, then knowing the zeros of $F_n(p)$ makes it
possible to reconstruct the entire function.

Another example of the fractal behavior is the frequency dependence of the
impedance of a large, hierarchically constructed circuit in the case of
small, real losses. In Ref. 4 it was found that in a random medium
consisting of capacitors, inductors and resistors with small resistance,
the local electric field is subject to anomalously strong fluctuations.
For example, in a 2D Dykhne medium consisting of two metals the imaginary
pert of the effective permittivity at a frequency between their plasma
frequencies is nonzero in the abscence of local absorption. This causes
the local electric field to diverge. We shall show that anomalously strong
fluctuations in the properties occur even in a regular hierarchical chain.
Strictly speaking in the abscence of absorption even self-averaging of the
system does not occur.

Consider now the electrical circuits in Fig.2. Let the initial element $i$
be a capacitor $C$ with impedance $Z_i= i/\omega C$ at frequency
$\omega$ and the impedance of the initial element of the iteration
sequence $Z_0=-iL\omega +R$, where $L$ is the inductance and $R$ is
the resistance.  Switching to dimensionless variables
$z_n=\omega C Z_n$, $z_0=-iLC\omega^2+RC\omega$, we obtain for the
circuit in Fig.2a

\begin{eqnarray}
\label{4}
z_{n+1}=i\frac{(2z_n+i)}{(z_{n}+i)}.
\end{eqnarray}

The explicit formula for $z_n$ has the form
$z_{n}=(if_{n+1}z_{0}- f_n)/(f_n z_{0}+ i f_{n-1})$,
where $f_n$ are the Fibonacci numbers

$$
f_n=\frac{1}{\sqrt{5}}\biggl[\bigl(\frac{1+\sqrt{5}}{2}\bigr)^{n+1}-
\bigl(\frac{1-\sqrt{5}}{2}\bigr)^{n+1}\biggr]
$$

The set of both poles and zeros has a limit point $-i(\sqrt{5} -1)/2$.
Near this point $z_{n}$ converges poorly.

In the example analyzed above, the final formula reduces to a linear
fractional transformation. In the general case the map (2) performed $n$
times leads to rational function of high order. The result of this mapping
in the space $z_0$ can be predicted if the stationary points and poles of
transformation are known. As a degree of the transformation increase, the
number of poles increase as $2^n$ and the convergence of the mapping
therefore becomes a fractal function of the initial impedance.

Fig.2b corresponds to the map

\begin{eqnarray}
\label{5}
z_{n+1}=\frac {z_n(z_{n}+i)}{2z_n+i}.
\end{eqnarray}

Figure 3 displays the impedance after 10 iterations as a function of
$z_0$. It is evident from the figure that in the abscense of absorption
the impedance is a fractal curve. According to the definition of
$z_0$, the fractal nature is reproduced in the frequency dependence
of the impedance. The fractal behavior is explained by the fact
that random circuits which resonate at the frequency $\omega$ arise
in the system. In the limit of small $R$, the width of a resonance
is determined by the quantity $R/L$. The total number of resonances
distributed in a characteristic frequency band of the order of
$\sqrt{LC}$ is determined by the number $2^n+1$ of the element in
the system. As the static conductivity of elements increases, the
resonances cease to overlap if $\sqrt{LC}2^{-n}L/R>1$, and the
absorption is smoothed out.

An important property of a percolation system is the percolation radius
	$r_c$. The geometric percolation radius is determined as the scale
	on which the probability of two sites belonging to the same
	cluster starts to decrease more rapidly than in power law. The
	correlation radius becomes infinite at percolation threshold.

 In  the analysis of the conductivity of a two-phase random system containing
phases with finite dc conductivities, it assumed that the correlation
radius is related to the ratio $h=\sigma_1/\sigma_2$ of the conductivities:
$r_c\sim h^{-\beta}$. However, the question of to determines the correlation radius
remains open in the case of an ac field. A necessarily real quantity, which
$r_c$ is, cannot be an analytic function of the impedances. Therefore the
function $r_c(\sigma_1,\sigma_2)$ cannot be suggested as an alternative.

In the model corresponding to Fig. 2b, the impedance approaches zero as the
number of iterations increases. In the limit of a sufficiently large number
of steps, the renormalization-group equation can be converted into the
differential equation $dz/dn=-z^2$ with the solution $z=1/(n+A)$.
It is natural to take as the correlation length, is units of
$n$, the value of n for which $z$ reaches this asymptotic law. For
real $z$, this occurs for $n\sim \ln z$.

 In Fig. 4, the number of iterations at which the impedance drops below 0.01
in modulus is displayed as a function of the initial impedance. One can see
how this dependence becomes fractal as R decreases.

 The correlation radius estimated from the smoothing of the resonances has the
form $n\sim \ln{(1/R)}/\ln{2}$. The fact that $r_c$ becomes
infinite can be understood if the system under study is interpreted
in the limit $R\to 0$ as a system of coupled, localized states for
a photon.  In such a formulation, the spectrum of the system
consists of separate lines, which do not vanish in the absence of a
relaxational interruption of the phases. At the same time, the substitution
$z\to i z$ gives the system coupled positive and negative
resistances, i.e., the system contains amplifying  sections, which
explains the absence of a thermodynamic limit.

  This work was partly supported by the Russian Fund for Fundamental Research,
Grants Nos. 95-02-04432 and 96-02-19353 and the Volkswafen Foundation.

\medskip

{\bf Picture signs}

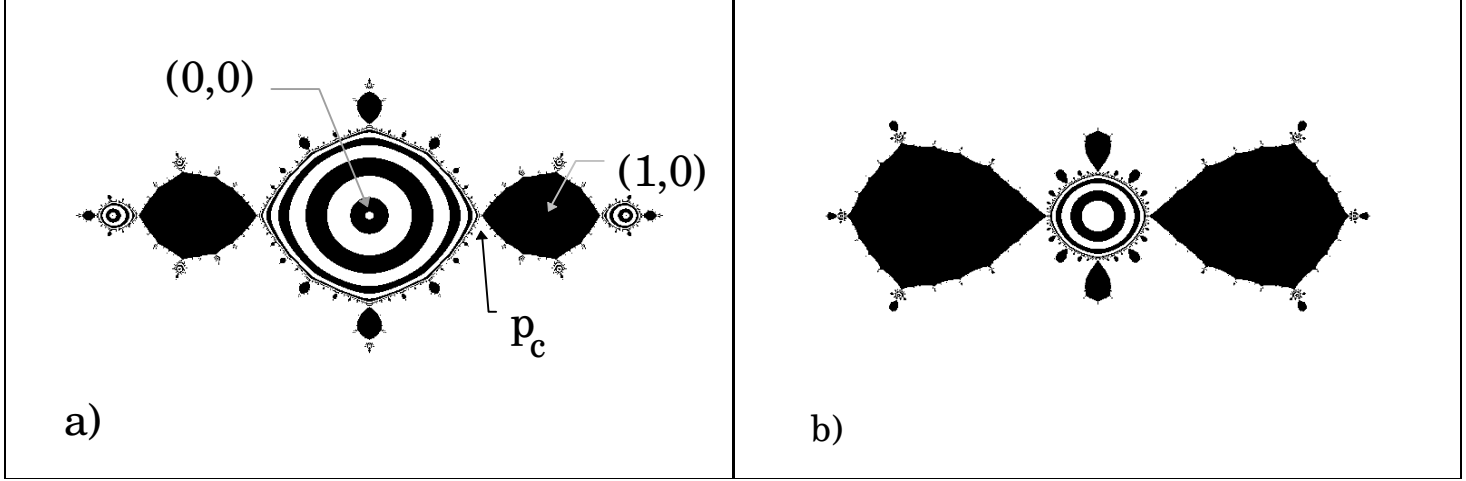
FIG. 1. a-Basin of attraction of the fixed point 0 corresponding to
nonpercolation with $d=2$. The regions where the number of steps required to
reach a neighborhood of the point 0 is even are colored in black, as are the
basins of attraction of other, nonfixed points, except infinity. b-Same as
Fig. 1a,$d=3$.

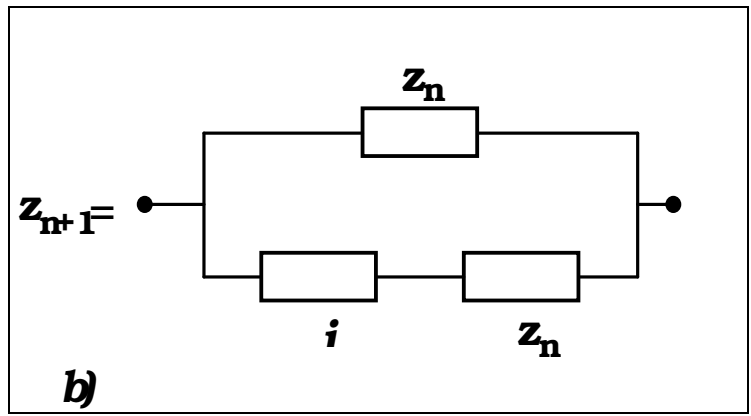
FIG. 2. Circuits of the hierarchical chains a and b corresponding to the
expressions (4) and (5).

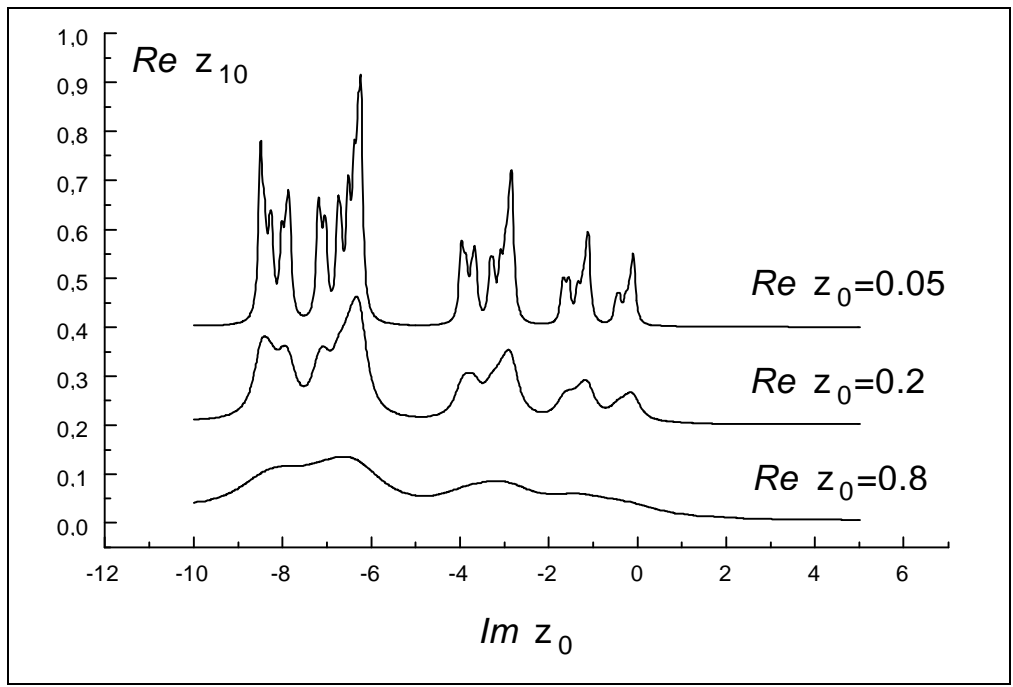
FIG. 3. Impedance of the chain in Fig . 2b as a function of the imaginary part
of the initial impedance after 10 iterations. Adjacent curves, except for the
bottom curve, are raised by 0.2 units relative to one another for better
resolution.

FIG. 4. a-Number of steps required to reach a neighborhood of $3 10^{-2}$ as a
function of the initial impedance for the circuit in Fig. 2b. Adjacent curves,
except for the bottom curve, are raised by 25 units relative to one another.
b-Dual coloring, similar  to Fig. 1, of the plane of initial impedances.

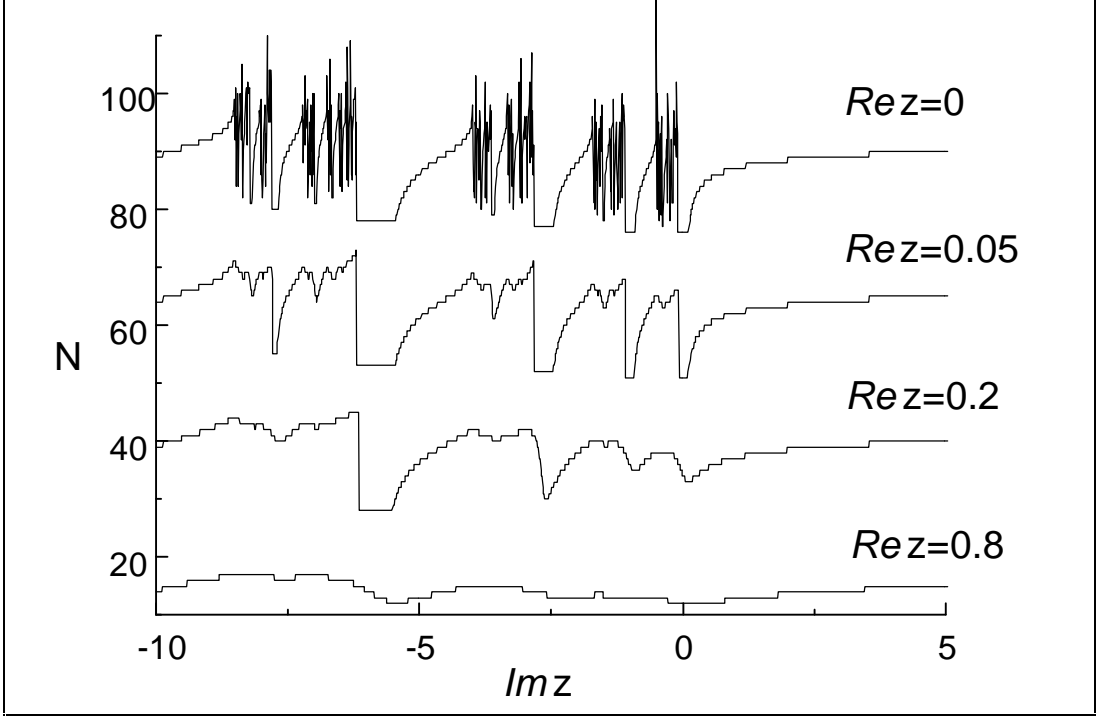
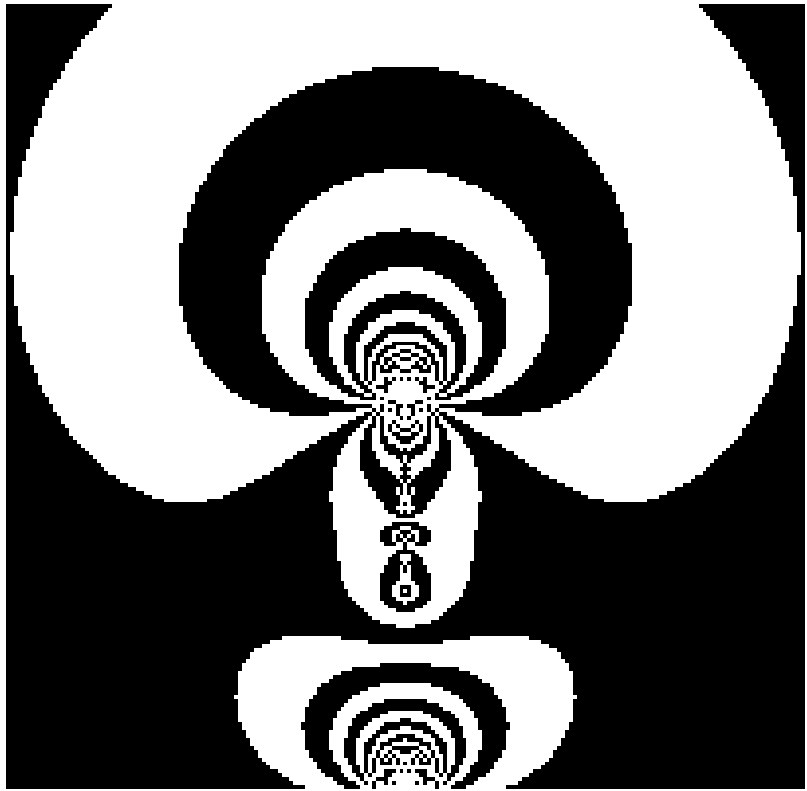
\end{document}